\documentclass[%
 reprint,
 superscriptaddress,
 amsmath,amssymb,
 aps,
 apl,
]{revtex4-1}

\usepackage{float}
\usepackage{subcaption}
\restylefloat{figure}
\usepackage{graphicx}
\usepackage{dcolumn}
\usepackage{bm}
\usepackage{inputenc}
\usepackage{multirow}
\usepackage{upgreek} 
\usepackage{hyperref}

\hypersetup{
colorlinks=true,
linkcolor=blue,
citecolor=blue,
urlcolor=blue
}

\usepackage{listings}

\graphicspath{{./images/}}

\usepackage{etoolbox}

\makeatletter
\patchcmd{\frontmatter@abstract@produce}
  {\vskip200\p@\@plus1fil
   \penalty-200\relax
   \vskip-200\p@\@plus-1fil}
  {}
  {}
  {}
\makeatother

\begin{document}

\title{TribChem: a Software for the First-principles, High-Throughput Study of Solid Interfaces and their Tribological properties.}

\author{Gabriele Losi}
    \affiliation{Department of Physics and Astronomy, University of Bologna, 40127 Bologna, Italy}
\author{Omar Chehaimi}
    \affiliation{Department of Physics and Astronomy, University of Bologna, 40127 Bologna, Italy}
\author{M. Clelia Righi}
    \email[]{clelia.righi@unibo.it}
    \affiliation{Department of Physics and Astronomy, University of Bologna, 40127 Bologna, Italy}

\begin{abstract}
High throughput first-principles calculations, based on solving the quantum mechanical many-body problem for hundreds of materials in parallel, have been successfully applied to advance many materials-based technologies, from batteries to hydrogen storage. However, this approach has not yet been adopted to systematically study solid-solid interfaces and their tribological properties. To this aim, we developed TribChem, an advanced software based on the FireWorks platform, which is here presented. TribChem is constructed in a modular way, allowing for the separate calculation of bulk, surface, and interface properties. At present the calculated interfacial properties include adhesion, shear strength, and charge redistribution. Further properties can be easily added due to the general structure of the main workflow.
\end{abstract}


\maketitle


\section*{\label{sec:intro}Introduction}

Integrating the experiments with computational tools and digital data is considered a key strategy to reduce the time and costs for materials discovery and deployment. In this context, first-principles high throughput calculations, which allow for the density functional theory (DFT) description of many materials in parallel, and in an automatized way, represent very powerful tools ~\cite{Curtarolo-2013,Haastrup-2018,Li_npj_comp_mat2022,Rosen_npj_comp_mat_2022,Hebnes_npj_comp_mat_2022,Choudhary_npj_comp_mat2022}. 
The calculated properties, usually highly accurate, are stored in databases, and eventually analyzed with the aid of machine-learning algorithms, allowing for the identification of general trends and predictions. Moreover, raw data are also stored so that the calculation of further properties and rigorous validations are possible. 

High throughput calculations have been successfully applied to advance several materials-based technologies, including superconductivity~\cite{wines_2022}, catalysis \cite{tran_JCIM2018, rosen_JCompChem2019,Rosen_npj_comp_mat_2022}, high-entropy alloying \cite{rittiruam_scirep_2022}. However, in this quickly developing framework, a systematic study of solid-solid interfaces and their tribological properties has not been addressed yet. Most probably this is due to the inherent difficulties that this kind of system, composed of two different lattices matched together, poses and to the fact that the community of references (the tribology, metallurgy, and mechanical manufacturing communities) traditionally relies on classical macroscopic engineering models.

Here we present TribChem, a software designed to perform the high-throughput study of solid interfaces. The software is composed by three main units for the study of bulks, surfaces, and interfaces. It is entirely written in Python and uses different packages from the Materials Projects~\cite{Jain-2013}. TribChem is based on FireWorks~\cite{CPE:CPE3505},  an open-source code for defining, managing, and executing workflows. Complex workflows can be defined using Python, JSON, or YAML, and stored using MongoDB. For connecting to the Materials Project database a high-level interface class named NavigatorMP has been developed.
To perform the Density Functional Theory (DFT) calculations, TribChem relies on the Vienna Ab initio Simulation Package (VASP)~\cite{Kresse-93, Kresse-94, Kresse-96, Kresse-96-2}.

A first workflow developed by our group~\cite{Restuccia-2018}, based on the Aiida platform~\cite{aiida}, was used for studying the interfacial properties of homogeneous interfaces. 
TribChem, contains several technical advancements with respect to this initial workflow, related to the creation of a proper error handling and the possibility to store and retrieve the data through a pubblicly accessible database, and has been extended to perform the study of heterointerfaces, such as those forming grain boundaries and heterojunctions. Simulating two different surfaces in contact is computationally much more demanding than considering equivalent surfaces. A common cell should be, in fact, identified to accommodate the two different lattices with a reasonably small mismatch. This typically increases the size of the simulated system and its complexity. Indeed, we introduced a procedure, based on the surface symmetries, to calculate the potential energy surface (PES) that describes the interaction of the two surfaces in contact as a function of their relative lateral position~\cite{Restuccia-2018}.  
We have successfully employed Tribchem to systematically search for the optimal interface geometry and accurately determine adhesion energies of hundred metals relevant to technological applications. This allowed us to populate a database of accurate values, which can be used as input parameters for macroscopic models\cite{Restuccia-2023}.

The aim of the present paper is to provide a technical description of the structure, main features and an example of use of TribChem. The first three sessions are devoted to the technical description of the software infrastructure, workflows and database. Finally, the instructions for the workflow execution and an example of use are presented.

\section{Implementation and architecture}
\begin{figure}[htbp]
    \centering
    \includegraphics[width=\linewidth]{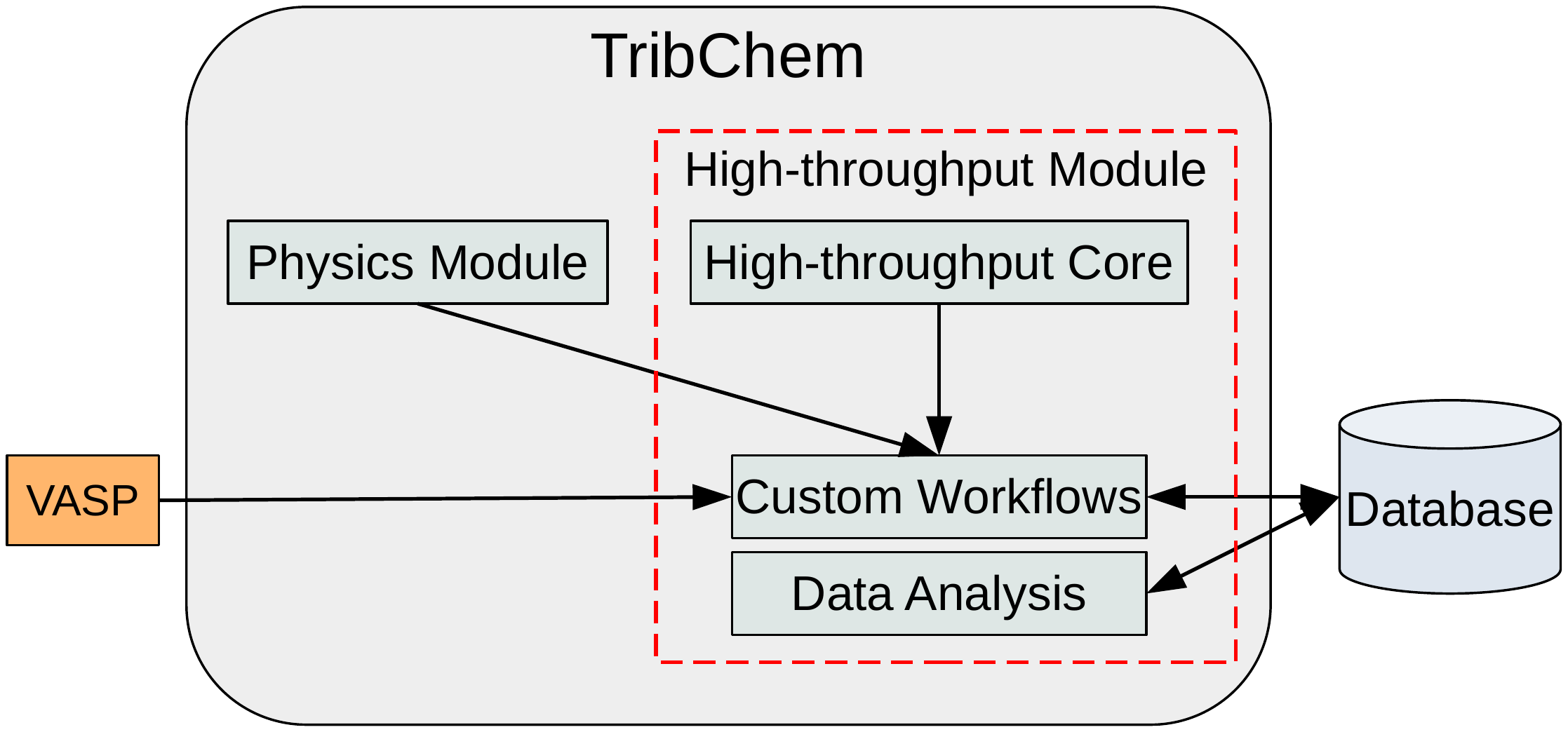}
    \caption{Schematic view of TribChem main components.}
    \label{fig:tribchem}
\end{figure}

The TribChem package consists of three main elements: the physics, the high-throughput modules, and the database (Fig.~\ref{fig:tribchem}).

\subsection{Physics module}
In the physics module, we implemented several functions to perform basic operations on solid-state physics, math, geometry, file manipulation, plotting, post-processing, and advanced operations on electronics, mechanical, and tribological properties. These functions are implemented in a highly modular way inside well-defined classes. Thanks to this design, it is possible to use these functions outside TribChem by importing them as external modules. The physics module relies on the Python Materials Genomics (Pymatgen)~\cite{pymatgen} package to manipulate bulk, slab, and interface structures and on the MPInterfaces~\cite{mpinterfaces} package, which implements the algorithm for geometrically matching two surfaces.

Each sub-module is briefly described below:
\begin{itemize}
    \item \verb!base!: It contains a collection of functions for manipulating atomic structures with periodic boundary conditions. It enables the replication of crystalline cells, the translation or rotation of atomic clusters, and the creation of an orthorhombic-shaped base area for any cell. It also includes mathematical functions for calculating gradients on three-dimensional surfaces and fitting data to polynomials of any degree. There is also a converter between the various physical units and CSV tables of the elemental materials classified based on their properties. Moreover, this module has advanced classes that allow working on the atomic structures of bulk, slab, and interface.
    \item \verb!chembas!: A class containing a set of functions useful for studying the adsorption of atoms on surfaces.
    \item \verb!dft!: I/O handlers for VASP, Quantum Espresso~\cite{qe}, and Lammps~\cite{LAMMPS} are available, allowing the user to load input files into versatile Python dictionaries for further processing or extract information of interest from output files.
    \item \verb!dynamics!: A set of tools for analyzing a dynamic simulation and calculating the positions, velocities, and forces acting on a single species or groups of atoms. It also includes many tools for plotting these properties.
    \item \verb!electronics!: A set of functions and classes helpful to combine the electronic charge densities of output files and plot the electronic structure's bands and density of states.
    \item \verb!ml!: An experimental sub-module containing some functions for the generation of datasets suitable to train machine learning models on them.
    \item \verb!tribology!: It includes tools for computing and analyzing the tribological properties of interfaces, such as functions for matching two surfaces, for calculating the Potential Energy Surface (PES), the Minimum Energy Path (MEP), the and shear strength by combining the high-symmetry points of the surfaces as described in previous papers of our group~\cite{hetero_workflow,Restuccia-2023}.
\end {itemize}
 
\subsection{High-throughput module}
We implement the core classes necessary to create the building blocks of the custom workflows in the high-throughput module. Thanks to the modularity creating new custom workflows to perform new high-throughput calculations, also outside the field of tribology, is greatly simplified. Moreover, inside this module, we created some functions and custom classes to perform data analysis of the results. Finally, we added all the functions necessary to communicate with the database, which is the central element of TribChem due to the workflow management, and we stored in it all the results.

One needs four different input files to run a VASP calculation: POTCAR, POSCAR, INCAR, and KPOINTS (for more information on the structure of these files, see the \href{https://www.vasp.at/wiki/index.php/The_VASP_Manual}{VASP Manual}). All these files are generated automatically by TribChem according to the calculation one would like to execute. The creation of the POSCAR and the POTCAR files is relatively easy. The pymatgen library generates the POSCAR by reading the pymatgen object from the database, while the same library also creates the POTCAR by reading the pseudopotential files saved in a location specified during the installation of TribChem. Instead, for generating the INCAR and the KPOINTS files, we created two classes, \verb!VaspInputSet! and \verb!MeshFromDensity!, that allow us to manipulate all possible setting options for file generation. Indeed, the INCAR and the KPOINTS files greatly differ from structure to structure. Hence, we developed custom classes that read from an external JSON file the parameters that should be passed as input.

\section{The FireWorks library}
We briefly introduce the logic of the FireWorks library to understand how the high-throughput module works. Every workflow written in FireWorks has three fundamental components: the FireTask, the FireWork, and the Workflow. The FireTask is the most basic component, which usually implements an indivisible task (e.g. launch a VASP calculation, or save data in the database). At the FireTask level, the input parameters are passed through the class variables \verb!required_params! and \verb!optional_params!. The FireWork has an intermediate level of complexity and can comprise one or more FireTasks, which are executed consecutively based on a given order (e.g. run a VASP calculation and then save the results in the database). Moreover, a FireWork can initialize and communicate with one or more new FireWorks (which in turn contain other FireTasks), generating a complex hierarchical structure where several operations are executed in a given order, sequentially or in parallel. Finally, the Workflow at the top level of the hierarchy, groups several FireWorks to execute a complex calculation (e.g. relax an entire structure or calculate the adhesion energy of an interface). The Workflow is the only unit which can be launched on a workstation.

In practice, a FireTask is a class characterized by two class variables for dealing with the input parameters: \verb!required_params!, \verb!optional_params!, and the method \verb!run_task! to run the instructions of the FireTask. In addition, the tasks done by the FireTask are usually implemented inside the class as methods and then called inside \verb!run_task! to be executed. Since all the FireTasks share common operations, we developed a general FireTask class named \verb!FireTaksTribChem!, in which we implemented all the methods common to the different FireTasks. All the custom FireTasks we developed inherit from the \verb!FireTaksTribChem! class, and thus having direct access to all its methods. In this way, the development of new FireTasks is greatly simplified and typically only necessitates the creation of specific methods. Of course, it is always possible to override the parent methods in the child classes, creating specific methods for the occurrence.

In short, these methods are: 
\begin{itemize} 
\item \verb!read_runtask!: Read the optional and required parameters of a FireTask. The user passes these as input arguments when initializing the FireTask object. 
\item \verb!read_params!: Update the dictionary parameters with default values read from a JSON file.
\item \verb!set_filter!: Create a filter to query the database to retrieve/store some data. 
\item \verb!is_done!: Check if a specific entry within a MongoDB~\cite{mongodb} document matches the filter. It is useful to check whether a calculation is already done and decide whether to run or stop a job. 
\item \verb!query_db!: Retrieve data from a MongoDB database. The database location and tables are specified with keywords: \verb!db_file!, database, collection, entry. 
\item \verb!update_db!: Update data in the matching field. Data are prepared to be stored in a database creating a dictionary according to the MongoDB guidelines. The database field to be updated is located by: \verb!db_file!, database, collection, entry. 
\item \verb!insert_db!: Add a new field to a database. The precise position is identified by \verb!db_file!, database, and collection. 
\item \verb!query_mp!: Query the Materials Project database, searching for a structure with a selected \verb!mp_id! or searching for the lowest energy structure given a specific chemical formula.
\item \verb!save_files!: Create the folders to save the output files on the local machine. 
\end{itemize}

\subsection{Workflow creation}
Thanks to the FireWorks library structure, we can employ several strategies to create complex workflows by combining FireTasks and FireWorks. During the development of TribChem, we decided to implement the FireTasks first and then create FireWorks by employing just one FireTask to maintain the highest level of modularity. We followed this approach with some exceptions. In some situations, it is helpful to have several operations inside a FireWork. For example, every time we run a VASP simulation, all the results must be saved in a proper location in the database. For this reason, we developed the function \verb!run_and_save! which returns a FireWork containing two FireTasks: one for running the simulation (\verb!FT_RunVaspSimulation!) and another one for saving the results (\verb!FT_MoveResults!).

This approach allowed us to exploit as much as possible the modularity given by the FireWorks library by creating atomic operations we used as building blocks to construct complex workflows.

The following example shows how a workflow is built. In this case, the workflow is composed of three serial FireWorks, each of them containing just one FiresTask:
\begin{center}
\centering
\begin{lstlisting}[language=bash]
ft1 = CustomFiretask(...)
ft2 = CustomFiretask(...)
ft3 = CustomFiretask(...)
fw1 = Fireworks(ft1, name="FireWork 1")
fw2 = Fireworks(ft2, name="FireWork 2")
fw3 = Fireworks(ft3, name="FireWork 3")
wf = Workflow([fw1, fw2, fw3], 
              links={fw1: fw2, fw2: fw3}, 
              name="Workflow")
\end{lstlisting}
\end{center}
Each FireWork communicates with the next one, and the execution order is specified inside the workflow definition as a dictionary in the variable called \verb!links!. Hence, the order is: \verb!fw1!, \verb!fw2!, and \verb!fw3!. 

\subsubsection*{The slab generator workflow}
\begin{figure}[htbp]
    \centering
    \includegraphics[width=\linewidth]{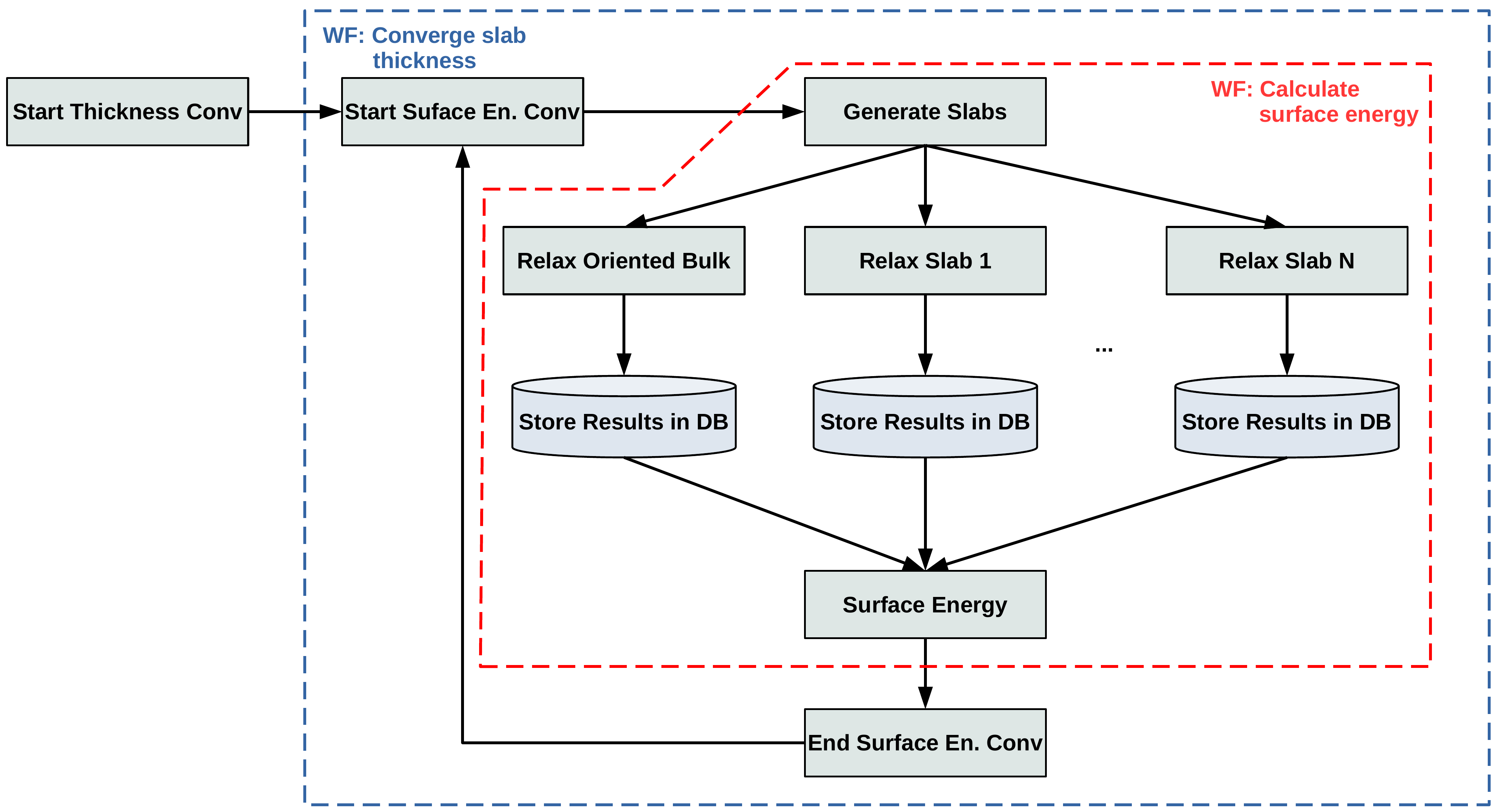}
    \caption{Flowchart of the two nested workﬂows used to generate crystalline slabs and converge their atomic thickness by evaluating the surface energy.}
    \label{fig:surfene}
\end{figure}

To better clarify the role of the modular approach we
used, in the following we describe, as an example, how the workflow
generates a slab with a given orientation by the convergence of its surface energy (schematically shown in Fig.~\ref{fig:surfene}).

\subsection{Initializing Firework}\mbox{} \\
The workflow that calculates the optimal thickness is composed of two workflows: the main workflow, which starts the surface energy convergence (\verb!Converge slab! \verb!thickness!), and the internal workflow, which calculates the surface energy (\verb!Calculate! \verb!surface energy!). As shown in Fig.~\ref{fig:surfene}, the internal workflow is called recursively until the convergence of the surface energy is reached. The \verb!Start Thickness Convergence! FireWork starts the whole workflow. During this step, the program checks first whether the requested slab has already been calculated. If it is not the case, the calculation is launched. In addition an initial check is done to control which slab thicknesses have already been calculated. The simulation is launched only for the systems that have not been computed before. These checks allow us to save time and computational resources.

\subsubsection*{WF: Converge slab thickness}\mbox{} \\
The \verb!Converge slab thickness! workflow consists of two FireWorks: 
\begin{itemize}
    \item \textbf{Start Surface Energy Convergence}: Calls the internal workflow which calculates the surface energy.
    \item \textbf{End Surface Energy Convergence}: Checks if the convergence has been reached. We check the convergence by comparing the computed surface energies with the reference value defined as the energy of the maximum allowed thickness. When the relative difference in surface energy of one slab is lower than a certain threshold, the convergence is reached, and the workflow finishes. If the convergence is not reached after the recursive calls, the convergence criterion is not satisfied, and the maximum thickness selected by the user is assumed to be optimized.
\end{itemize}

\subsubsection*{WF: Calculate the surface energy}\mbox{} \\
This internal workflow calculates the surface energy and is composed of four different FireTasks:
\begin{itemize}
    \item \textbf{Generate Slabs}: Generates a slab or a list of slabs starting from a bulk structure. The slab has the desired orientation, specified by the Miller indexes, thickness, and vacuum to be included in the supercell for separating it from its replicas.
    \item \textbf{Relax Structure}: the geometry optimization is performed using VASP.
    \item \textbf{Store Results}: Saves the results of the first principle simulations in the database.
    \item \textbf{Surface Energy}: Calculates the surface energy and saves the result in the database. 
\end{itemize}

\subsection{Workflows available in TribChem}
The currently available workflows of the TribChem package are:
\begin{itemize}
    \item Query online databases to save data to an internal database.
    \item Kinetic energy cutoff convergence.
    \item K-points convergence.
    \item Cohesion energy calculation.
    \item Calculation of the optimal thickness for a slab.
    \item Surface energy calculation.
    \item Matching and generation of a solid interface.
    \item Calculation of the PES.
    \item Evaluation of the adhesion energy.
    \item Extraction of the MEP and shear strength.
    \item Calculation of the Perpendicular Potential Energy Surface (PPES).
    \item Charge displacement analysis.
    \item Adsorption of atoms on surfaces.
\end{itemize}

\section{Database}
The database used by the FireWorks library is MongoDB~\cite{mongodb}, a NoSQL database which uses JSON like-documents to store and manipulate data. When a new installation of TribChem is carried out, the user creates several databases. The most important and relevant ones are the \verb!FireWorks! database, in which all the data for the workflow execution are saved as well as the simulation results, and some relevant metadata such as the location of the VASP output files and the \verb!tribchem! database. We designed this custom database to save, retrieve, and make it easier to share the results of the high-throughput calculations with the scientific community.

When performing a high-throughput study, it is critical to have powerful tools to efficiently perform the Input/Output (I/O) operations. Indeed, the amount of data generated is very high, and it is necessary to have the proper tools to manage them. To this end, we created several sub-modules containing Python tools for efficiently performing I/O operations. The class \verb!Navigator! combines several low-level functions of the PyMongo~\cite{pymongo} library. Starting from this class, we developed powerful functions to automatize the query operations inside all the FireTasks of TribChem. Technically, we did this by implementing the functions as methods of the main FireTask \verb!FiretaskTribchem!, which is used by all the FireTasks of TribChem. This design greatly simplifies the structure of the software since it hides the implementations of all the functions in just one class, making them more maintainable, less error-prone, and usable also by non-technical users.

We also developed a high-level interface class named \verb!NavigatorMP! for connecting to the Materials Project database.

\subsection*{The tribchem database}
\label{sec:database_design}
\begin{figure}[htbp]
    \centering
    \includegraphics[width=\linewidth]{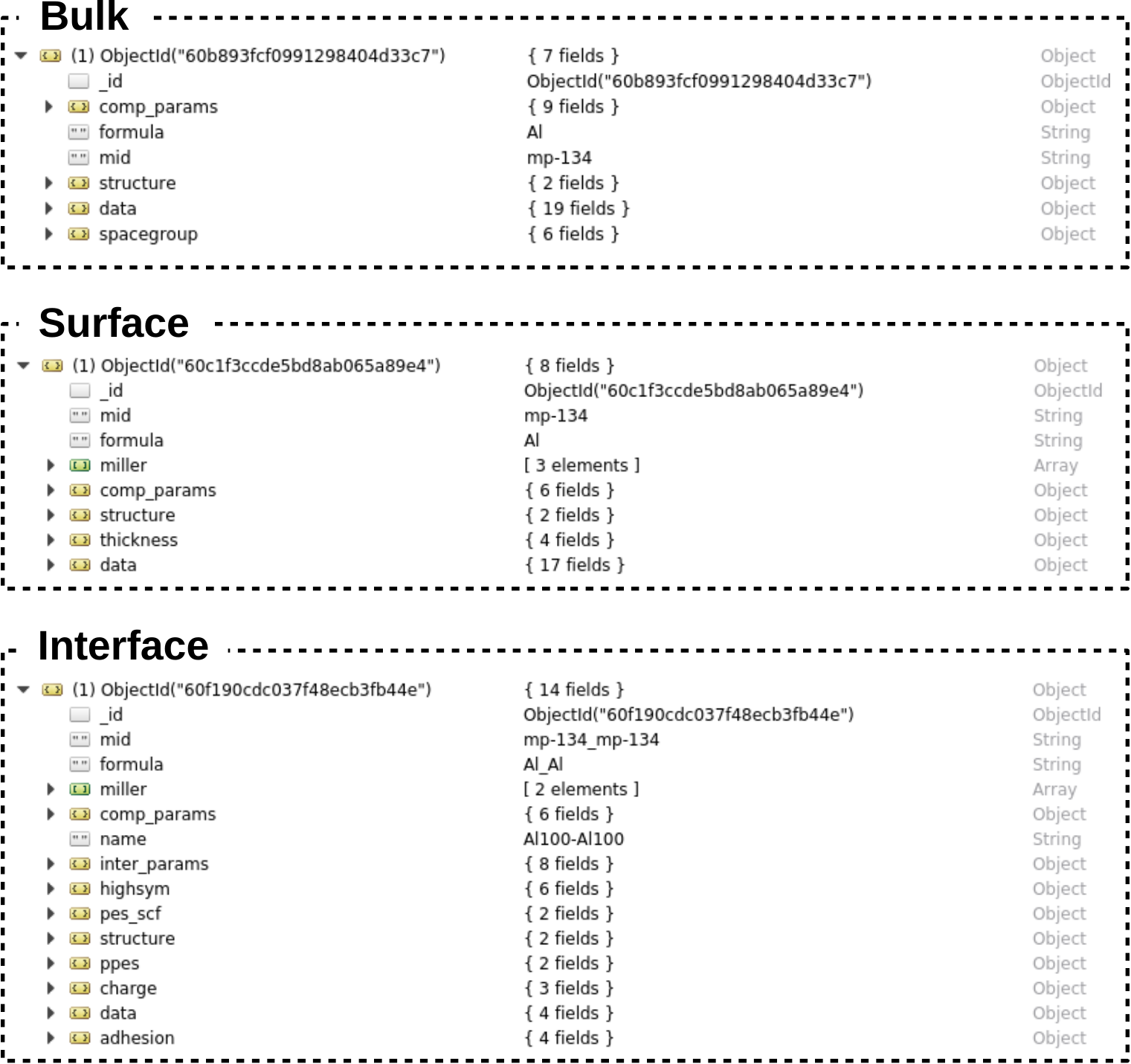}
    \caption{Typical design of the MongoDB documents in our database, showing how bulks, surfaces, and interfaces entries are saved. Aluminum is used as a case example.}
    \label{fig:db_structure_al}
\end{figure}

We carefully designed the structure of the \verb!tribchem! database to assure the best usability, maintainability, and robustness of the data stored inside it. This database includes several collections divided into different structures and functionals used for performing the DFT calculations. In our case, we have three main collections classes about bulks, slabs, and interfaces. The notation we used to name these collections is \verb!functional.material!. For example, the collection of the elemental bulks calculated with the Perdew-Burke-Ernzerhof (PBE) functional is named \verb!PBE.bulk_elements!. The structure of the three main collections is shown in Fig.~\ref{fig:db_structure_al}. 

The main problem when working with data stored in a database is uniquely identifying each element. For this reason, we defined a set of identifiers for each of the three main collections we created. 
 
 In the case of the bulk, the main identifier is the material identifier (\verb!mid!), which usually corresponds to the Materials Project ID for a given material, but it could be any alphanumeric value. In the case of aluminium, \verb!mid! corresponds to mp-134 (Fig.~\ref{fig:db_structure_al}). We added two optional identifiers to the main one: \verb!formula! and \verb!name!. The former usually corresponds to the chemical formula of the material, and the latter is used as an additional identifier useful to speed up the search and avoid misinterpretations in some cases. For example, in the case both diamond and graphene structures are saved in the bulk collection, setting \verb!name! to the structure name would avoid the conflict of having different elements with the same identifiers.

For identifying a slab, in addition to \verb!mid! and \verb!formula! identifiers, the crystalline orientation, defined by the Miller index, is used. The \verb!miller! identifier is a Python list in the form [\verb!h!, \verb!k!, \verb!l!] (e.g. [\verb!1!, \verb!1!, \verb!1!]).

Finally, we defined the identifiers for an interface. Their name is the same for the bulk and slab collections, but their values are made up by combining the identifiers of the two slabs forming the interface. The resulting notation is: \verb!mid=mid1_mid2!, \verb!formula=f1_f2!, \verb!miller=[[h1, k,1 l1],! \verb![h2, k2, l2]]!, and \verb!name=f1m1-f2m2!. In this case, the \verb!miller! identifier is a list containing two lists (each for the orientation of the slab components), and \verb!name! is the formula combination of the two materials with the string version of the Miller indices of the relative slab. For example, a homogeneous interface of aluminum is identified by (Fig.~\ref{fig:db_structure_al}):

\begin{center}
\centering
\begin{lstlisting}[language=bash]
mid = "mp-134_mp-134"
formula = "Al_Al"
miller = [[1, 0, 0], [1, 0, 0]]
name = "Al100-Al100"
\end{lstlisting}
\end{center}

In addition to the consistent definition of identifiers, we also defined a coherent structure for saving the inputs, outputs, and computational parameters for running the VASP simulations. In every document the fields \verb!structure!, \verb!data!, and \verb!comp_params! can be found. 

The field \verb!structure! includes two other fields: \verb!init!, which contains the primitive structure for the bulk downloaded from the Materials Project database, and for the slab and interface created by our workflow; and \verb!opt!, which contains the optimal structure obtained after the relaxation. Both these structures are saved as pymatgen objects.

The field \verb!data! contains the results of the VASP calculations, such as the total energy, the forces and stresses, bandgap, and many others. All the files produced by VASP are stored in the \verb!calculations! folder, which is generated automatically in the same location as the TibChem installation folder that keeps the same hierarchical structure of the collections.

The field \verb!comp_params! contains the computational parameters used by VASP. In this field, one can find the K-point density, the cutoff energy, whether the material is a metal or not and many others. If this field does not sufficiently define all the computational parameters, the user can introduce new ones. For example, we need information about the matching during interface formation. For this reason, we created the field \verb!inter_params!.

Finally, other fields can be created to save the results of specific calculations. In the case of the interface, where there are many kinds of calculations and additional properties can be calculated, there are additional fields, such as \verb!highsym!, \verb!pes!, \verb!adhesion!, \verb!ppes!, and \verb!charge!. These fields can store data conserning the high symmetry points, the energies of the PES, the adhesion energy, the perpendicular potential energy surface (PPES), and the charge displacement calculation, respectively (Fig.~\ref{fig:db_structure_al}).

\section{Execution of a workflow}
\begin{figure*}[htbp]
    \centering
    \includegraphics[width=0.8\linewidth]{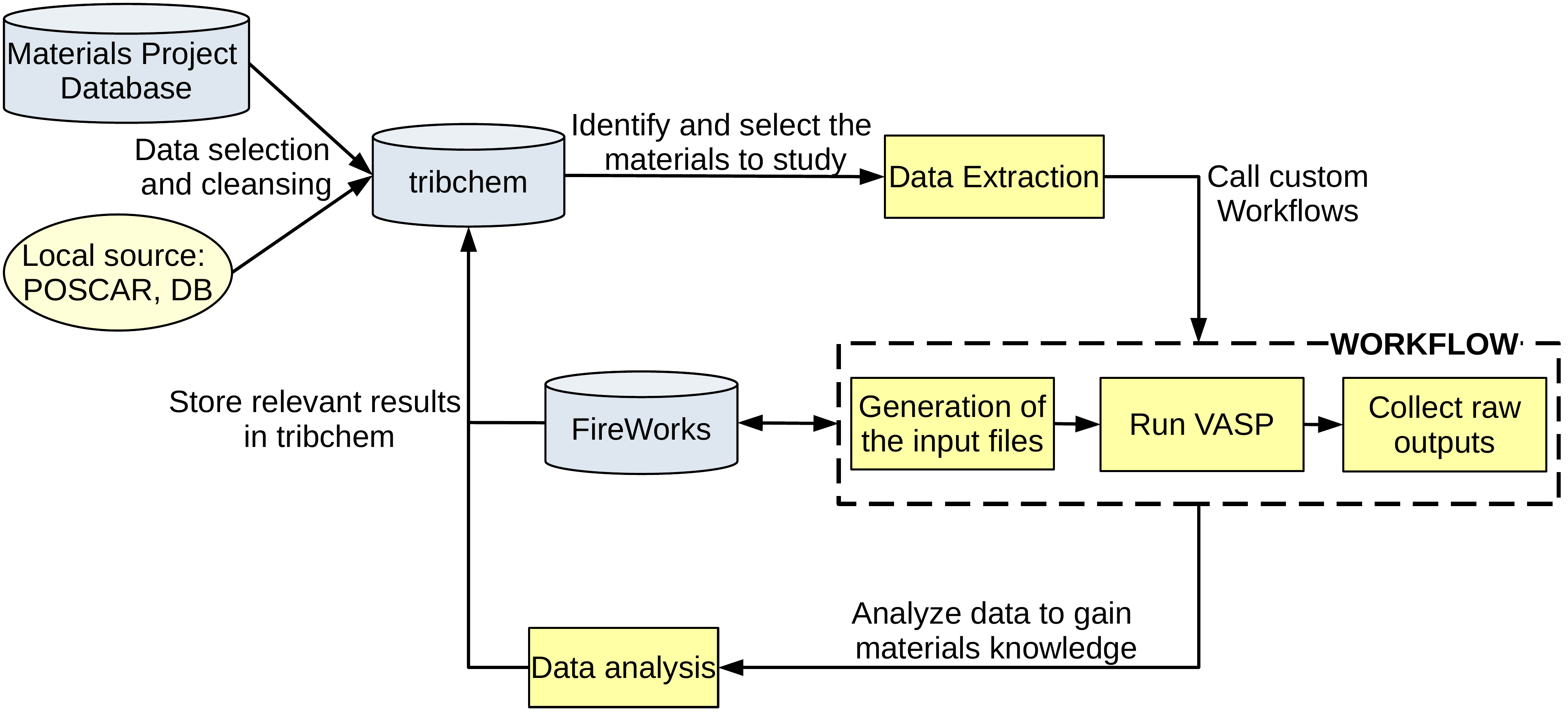}
    \caption{Flowchart showing the main logic in the execution of a workflow within the TribChem package. Four sequential logical steps are of major importance:
    1. Recovery of materials data, 2. Execution of a custom workflow and results collection, 3. Data analysis on the output data.}
    \label{fig:db_flow}
\end{figure*}

This section will explain the logic behind executing our high-throughput workflows. As shown in Fig.~\ref{fig:db_flow}, different components are involved during the execution of a workflow, with a continuous flow of data and instructions. 

The first step in running a workflow consists in the selection of the desired material with its crystallographic information. The primitive structures can be stored in the \verb!tribchem! database by downloading them from the Materials Project website, a local database, or directly from input files. Currently, only the Materials Project database can be used as an online resource. However, it is possible to add in the future the application programming interface (API) directives for the new database inside the TribChem project. Indeed, these new functions would write the new data into the unchanged \verb!tribchem! database, hence the same structure and functions remain the same. 

Once the crystalline structures are loaded into \verb!tribchem!, the workflows of interests can be executed. The execution of a workflow occurs in two steps: the first is to write the workflow in the database, and the second is to run the calculations (for more information on how this can be done in practice, see the user manual of TribChem~\cite{tribchem_wiki}). Based on the high-throughput study the user would like to perform, several workflows can be generated and written in the database and then run. Based on the experience we gained in using TribChem, we suggest running the same type of workflow to avoid possible race conditions. For example, the workflow that generates the slab is much faster than the workflow that calculates the PES and the adhesion energy. Although it is possible to run these two workflows at the same time for many different materials and interfaces, what would happen is that the faster workflows would be blocked by the slower ones, slowing down the whole process of getting new results. 

Our workflows are designed to be also used outside a high-throughput procedure. Indeed, they can take as input an atomic structure directly from a POSCAR file by loading it as a pymatgen object. In this way, it is also possible to use TribChem in simple scripts.

Once the workflows are completed, TribChem provides several data analysis tools. 

For more information on the practical usage of TribChem, see the user manual~\cite{tribchem_wiki}.

\subsection{Quality control}
We carried out the quality control of TribChem by using a set of dedicated functions present in the test folders of each module and submodule. The users can find more information on how to check if the installation process was successful and if the main functionalities are correctly working in the user manual~\cite{tribchem_wiki}.

\subsection{Availability}

\subsubsection*{Operating system}
The Linux operating system is required for the execution of TribChem. It is possible to use TribChem on the most relevant operating systems (Windows, macOS, and Linux) to create the workflows and analyse the data. However, we suggest using Linux.

\subsubsection*{Programming language}
Python 3.8.8.

\subsubsection*{Additional system requirements}
There are no particular requirements for using TribChem. The DFT calculations can usually be run on a high-performance computing system (HPC) to maximise the computational efficiency, while the workflow creation and data analysis can also be done on an ordinary desktop system.

\subsubsection*{Dependencies}
TribChem uses the following dependencies. Besides VASP~\cite{Kresse-93, Kresse-94, Kresse-96, Kresse-96-2}, which is a proprietary software and must be purchased from the official channels (see \\ \href{https://www.vasp.at/}{https://www.vasp.at/}), all the other packages can be installed with conda or pip (see the installation guide of TribChem).

\begin{itemize}
    \item atomate$\geq$0.9.9~\cite{atomate} A Python library containing a collection of FireWorks workflows which make easy to perform complex materials science computations.
    \item ASE$\geq$3.21.1~\cite{ase} The Atomic Simulation Environment (ASE) is a Python library for manipulating, running, visualizing and analyzing atomistic simulations.
    \item FireWorks$\geq$1.9.7~\cite{fireworks} A Python library used for defining, managing, and executing workflows in computational materials science.
    \item Matplotlib$\geq$3.3.4~\cite{matplotlib} A graphical library for creating data visualization in Python.
    \item MongoDB$\geq$4.0.3~\cite{matplotlib} The Python driver for the MongoDB database.
    \item MPinterfaces$\geq$2020.6.19~\cite{mpinterfaces} A Python library that enables high throughput DFT analysis of arbitrary material interfaces.
    \item NumPy$\geq$1.20.2~\cite{numpy} A high-level scientific computing library.
    \item pymatgen$\geq$2022.0.8~\cite{pymatgen} The Python Materials Genomics is a Python library for materials analysis.
    \item SciPy$\geq$1.6.2~\cite{scipy} A scientific computing library containing advanced algorithms on optimization, fitting, and other classes of numerical problems.
    \item VASP$\geq$6.2.0~\cite{Kresse-93, Kresse-94, Kresse-96} The Vienna Ab initio Simulation Package (VASP) is a software which enables to perform DFT calculations.
\end{itemize}

\section{Example of TribChem use}

In the following section we show how Tribchem can be used to compute in an automatic and systematic fashion 
the adhesion energies of the K(110) surface interfaced with a set of 17 technologically relevant metal surfaces.   
All the main workflow steps will be described starting from the bulk creation, the slab thickness optimization 
and surface energy calculations, and ending with the heterointerfaces creation and computation of the adhesion energy.
This example can be considered a furter guide for using TribChem to study hetero interfaces and tribological properties.

\subsection*{Bulk calculations}

The first step in the study of any interface is bulk optimization: TribChem identifies the bulk structure equilibrium geometry 
and optimizes the computational parameters for the 
selected elements. In order to do so, the program selects the optimal kinetic energy cutoff 
by converging, with respect to this computational parameter, the bulk modulus and the cell volume. 
To perform this task and launch the calculations, the user has to execute the following command within a Python virtual environment (here applied to the case of K bulk):

\begin{lstlisting}[language=bash]
    tribchem workflow converge_bulk 
    mids="[mp-58]" formulas="[K]"
\end{lstlisting}

More in details: a Birch-Murnaghan equation of state~\cite{Birch-1947} is used to fit the ab-initio data for increasing cutoffs.
The bulk modulus and the cell volume are extrapolated, when the relative difference between two consecutive steps is below 1\% for both 
quantities, convergence is reached.
As an example, data obtained for K are reported in Table~\ref{tab:K_cutoff}.

\begin{table}[htbp]
    \centering
    \caption{Data of the bulk convergence for K. The energy cutoff in eV, the bulk modulus in GPa, the cell volume in \r{A}$^3$ and the lattice parameter in \r{A} are reported. The bold values are the ones identified as converged.}
    \label{tab:K_cutoff}
\begin{tabular}{c c c c}
Energy cutoff & Bulk modulus & Cell volume & Lattice parameter \\
\hline
200 & 3.218 & 74.477 & 5.301 \\
225 & 3.551 & 73.858 & 5.286 \\
250 & 3.491 & 73.523 & 5.278 \\
\textbf{275} & \textbf{3.482} & \textbf{73.630} & \textbf{5.281} \\
300 & 3.488 & 73.668 & 5.282 \\
325 & 3.490 & 73.677 & 5.282 \\
\hline
\end{tabular}
\end{table}

In bold, the converged value of the bulk modulus and lattice parameter and the optimal kinetic energy cutoff are highlighted. 
The computed values are remarkably close to the experimental values ~\cite{Liu-1986} (3.0 GPa for the bulk modulus and 5.328 \r{A} for the lattice 
parameter), confirming the reliability of DFT and of our workflow in the determination of the this bulk properties.

With the same \texttt{tribchem} command, it is also possible to identify the optimal k-point sampling. In this case, TribChem 
fixes the energy cutoff at the optimal value while varying the k-point density until convergence is reached. For K, an optimal k-point density of 5.4 \r{A}$^{-1}$ is identified.

\subsection*{Slab and surface energy calculations}

The second step in the study of an interface is the generation and optimization of the two mating surfaces: as previously 
explained, TribChem creates a 
slab structure and then optimizes its thickness and computes the surface energy. To execute this workflow, the following 
command must be used (here we consider the K(110) surface):

\begin{lstlisting}[language=bash]
    tribchem workflow converge_slab
    mids="[mp-58]" formulas="[K]"
    miller="[[1, 1, 0]]"
    tmin=4 tmax=12
\end{lstlisting}

\texttt{tmin} and \texttt{tmax} identify the range of slab thicknesses to be considered (4 to 12 layers in this case). 
Specifically, TribChem computes the surface energy $E_{\gamma}$ and identifies the optimal slab thickness by fitting 
the slabs total energies $E_{slab}(N)$  at different layer thicknesses $N$ through the following formula, obtained
by inverting the surface energy definition:

\begin{equation}
	E_{slab} = A \cdot E_{\gamma} + N \cdot N_{at} \cdot \epsilon_{bulk}
\end{equation}

Where $N_{at}$ is the number of atoms per layer, $\epsilon_{bulk}$ is the bulk cohesive energy, $A$ is the slab in-plane area, and $N$ is the number of layers in the slab system. In this way, $E_{\gamma}$ is a fitting parameter, and, when convergence
is reached, it is returned as the surface energy of the system. For K(110), 4 layers are identified as the optimal slab 
thickness, correspondingly the surface energy is 0.11 J/m$^2$, which matches precisely the value reported 
by the Materials Project~\cite{Tran_scidata2016}. 

\subsection*{Homogeneous interface and PES calculation}
\begin{figure}[htbp]
    \centering
    \includegraphics[width=0.75\linewidth]{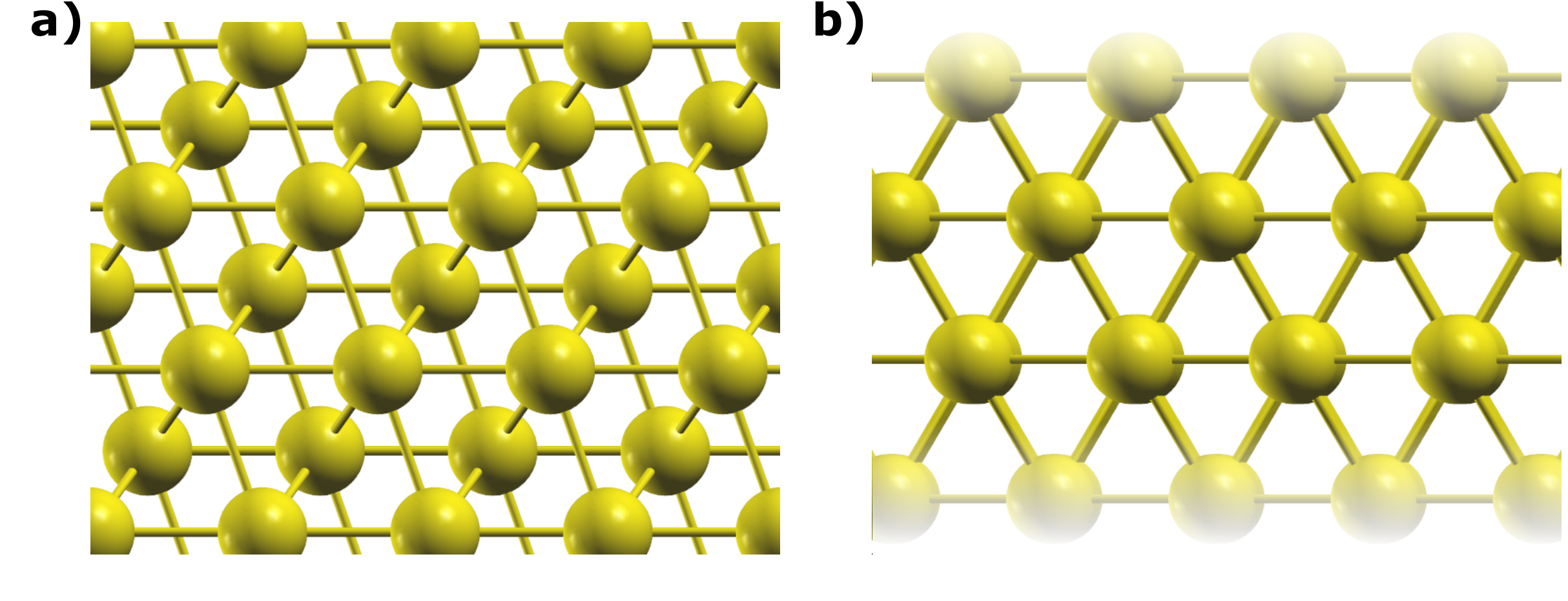}
    \caption{Top (panel a) and lateral (panel b) ball and stick representation of a K interface. The yellow color represents the K atoms.}
    \label{fig:K-K_view}
\end{figure}

Homogeneous interfaces are the simplest interfaces that can be studied as no cell matching is required. 
To generate and compute both the PES and adhesion energy of the K(110)/K(110) interface the following
command must be used: 

\begin{lstlisting}[language=bash]
    tribchem workflow
    calc_interface wf="[interface, pes, adhesion]"
    mids="[mp-58, mp-58]" formulas="[K, K]"
    miller="[[1, 1, 0], [1, 1, 0]]"
\end{lstlisting}

The \texttt{wf="[interface, pes, adhesion]"} command indicates that the user wants to launch the \texttt{interface}, \texttt{pes} and \texttt{adhesion} workflows altogether. First of all the \texttt{interface} workflow is launched and the interface structure of the required system is generated. A visual representation of the K-K interface is shown in Fig.~\ref{fig:K-K_view}.

Secondly the \texttt{pes} workflow begins: TribChem builds the PES landscape by mating the two slabs in different 
relative lateral positions and by collecting, for each displacement, the adhesion energy $E_{adh}$ given by:

\begin{equation}
    E_{adh} = \frac{E_{tot} - E_{top} - E_{bot}}{A}
\end{equation}

\begin{figure}[htbp]
    \centering
    \includegraphics[width=0.75\linewidth]{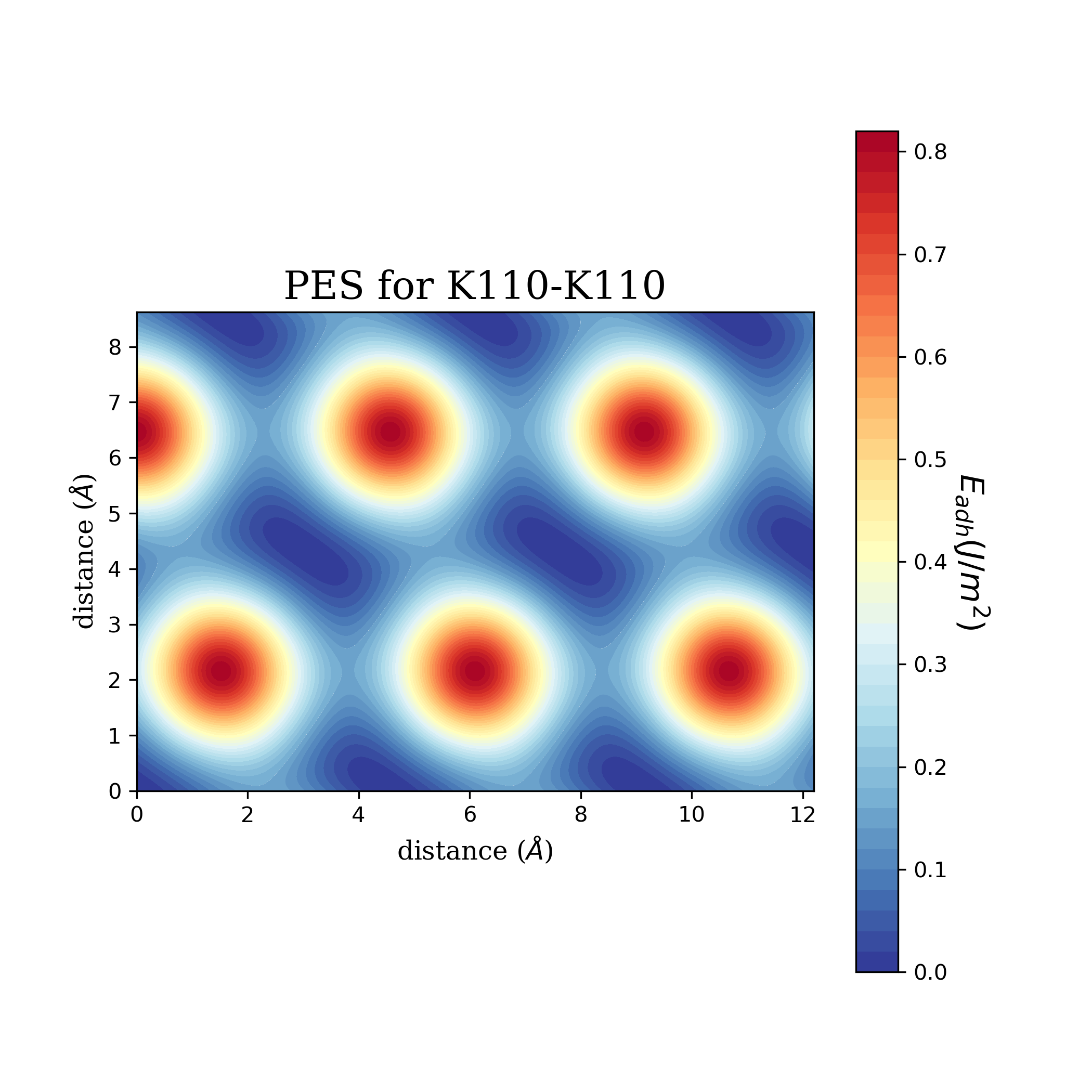}
    \caption{PES for the K homogeneous interface}
    \label{fig:K-K_pes}
\end{figure}

Where $E_{tot}$ is the energy of the interface, $E_{top}$ ($E_{bot}$) is the energy of the isolated top (bottom) slab composing the interface, and $A$ is the in-plane area. 
The relative lateral positions are identified by Tribchem by pairing the high symmetry points of the two surfaces. 
For the K(110) surface, four high symmetry points (the on-top, the hollow, the long bridge and the short bridge sites) 
are found, generating six non-equivalent lateral displacements for the K(110)/K(110) interface. 
Collecting the adhesion energy in every lateral displacement and interpolating them through the radial basis function makes it possible to obtain a 2D representation of the PES, as shown in Fig.~\ref{fig:K-K_pes}.

The final interface adhesion energy is the PES minimum, a value that is identified and stored in the database with the command \texttt{adhesion}. For the K(110)/K(110) interface, we obtain an adhesion energy of 0.20 J/m$^2$, almost twice the 
computed surface energy (0.11 J/m$^2$). This results validates the three interface workflows as, by definition of 
adhesion energy, its value in homostructure should be twice the surface energy. 

\subsection*{Heterogeneous interface with other metals}

\begin{figure}[htbp]
    \centering
    \includegraphics[width=0.8\linewidth]{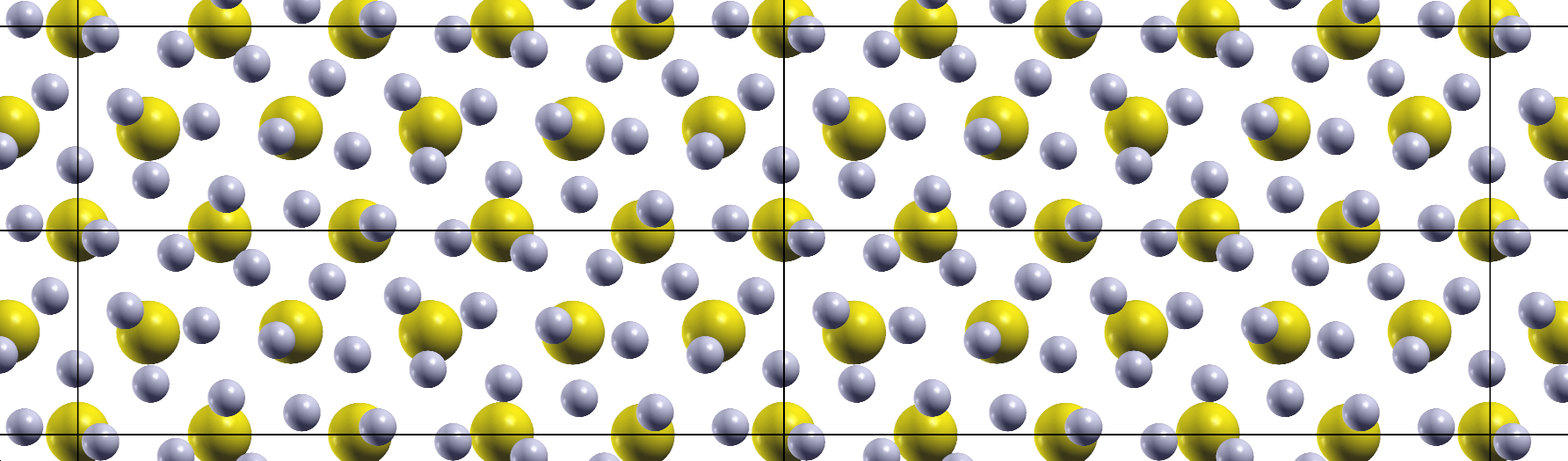}
    \caption{Top view of the interfacing atomic layers of the K/Al interface. The yellow (gray) balls represent the K (Al) atoms. The solid black lines are the boundaries of the supercells.}
    \label{fig:K-Al_match}
\end{figure}

Finally, we generated the heterostructures by mating the K(110) surface with the most stable faces of 17 different metals: 
Ag, Al, Au, Cr, Cu, Fe, Ir, Mg, Mo, Ni, Pt, Rh, Sc, Ti, V, W, and Zn. The surfaces are matched by MPInterface~\cite{mpinterfaces}, which makes use of the Zur alogirthm~\cite{Zur-1984}. 
Such a procedure allows to optimize the in plane supercell area within the allowed mismatch for the cell sides and angles. 
An example of the result of this  matching procedure is shown in Fig.~\ref{fig:K-Al_match} for the K(110)/Al(111) 
interface case.

\begin{figure}[htbp]
    \centering
    \includegraphics[width=\linewidth]{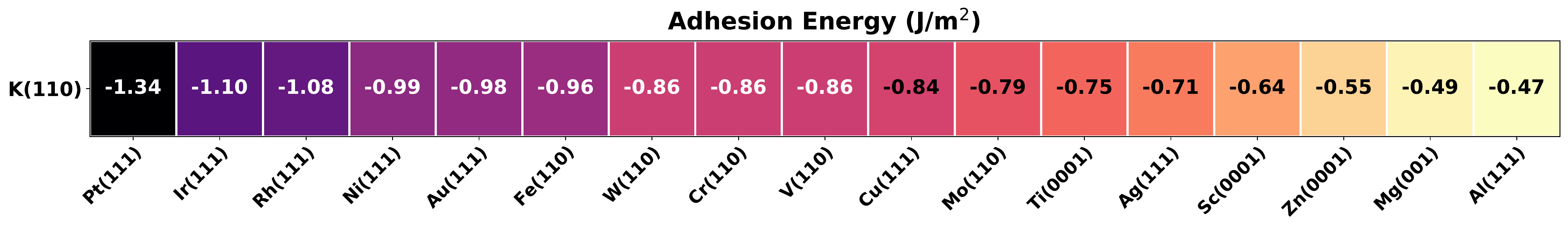}
    \caption{Adhesion energies data set for the heterogeneous interfaces obtained by combining K with 17 different metals. The data are sorted in ascending order.}
    \label{fig:adhesion}
\end{figure}

Using the same command shown for the homogeneous interface, TribChem generates the interfaces in different relative lateral positions to identify the energy minimum and compute the adhesion energy of the system. The results of these calculations are reported in Fig.~\ref{fig:adhesion}.

K has the largest adhesion with Pt, whereas the lowest interaction is with Al. In Ref.~\cite{Restuccia-2023} more than a hundred heterogeneous interfaces have been studied employing this workflow. 
The large amount of data, in that case, allowed for the use of a machine learning algorithm to determine predictive models
of adhesion energies in terms of properties of single slabs. 

\section*{Acknowledgements}
The development of TribChem is part of the ”Advancing Solid Inter-face and Lubricants by First Principles Material Design(SLIDE)”  project  that  has  received  funding  from  the European Research Council (ERC) under the European Union’s Horizon 2020 research and innovation program (Grant agreement No. 865633). The authors want to thanks Dr. Paolo Restuccia for providing the data of the example shown in the manuscript and for fruitful discussion.

\bibliography{references}

\end{document}